\documentclass[sigconf, nonacm]{acmart}

\usepackage{tikz}
\usetikzlibrary{arrows.meta,positioning,shapes.geometric,fit,backgrounds}

\definecolor{LaneA}{RGB}{230,243,255}  
\definecolor{LaneB}{RGB}{255,246,230}  
\definecolor{DataNode}{RGB}{69,123,157}
\definecolor{MLNode}{RGB}{231,111,81}

\tikzset{
  laneBG/.style={rounded corners, draw=black!25, fill=#1, inner sep=10pt},
  box/.style={rectangle, rounded corners, draw=black, line width=0.4pt,
              minimum width=48mm, minimum height=9mm, align=center, font=\footnotesize, fill=white},
  boxData/.style={box, draw=DataNode!85, fill=LaneA},
  boxML/.style={box, draw=MLNode!85, fill=LaneB},
  handoff/.style={box, very thick, draw=black, fill=gray!25},
  flow/.style={-Latex, line width=0.5pt},
  laneLabel/.style={font=\scriptsize\bfseries}
}

\AtBeginDocument{%
  }

\setcopyright{acmlicensed}
\copyrightyear{2026}
\acmYear{2026}
\acmDOI{XXXXXXX.XXXXXXX}
\acmConference[CAIN '26]{5th International Conference on AI Engineering – Software Engineering for AI}{April 12--18, 2026}{Rio de Janeiro, Brazil}
\acmISBN{978-1-4503-XXXX-X/2018/06}

\begin{document}

\newcommand{\gnote}[1]{\textcolor{red}{[[[#1]]]}}
\newcommand{\rnote}[1]{\textcolor{blue}{[[[#1]]]}}
\def\XXX{{\huge XXX}}


\title{From Pre-labeling to Production: Engineering Lessons from a Machine Learning Pipeline in the Public Sector}

\author{Ronivaldo Ferreira}
\affiliation{%
  \institution{Federal University of Pará}
  \city{Belém}
  \state{Pará}
  \country{Brazil}}
\email{ronivaldo.junior@icen.ufpa.br}

\author{Guilherme da Silva}
\affiliation{%
  \institution{Faculty of Gama, University of Brasília}
  \city{Brasília}
  \state{Distrito Federal}
  \country{Brazil}}
\email{guilherme.daniel@aluno.unb.br}

\author{Carla Rocha}
\affiliation{%
  \institution{Faculty of Gama, University of Brasília}
  \city{Brasília}
  \state{Distrito Federal}
  \country{Brazil}}
\email{caguiar@unb.br}

\author{Gustavo Pinto}
\affiliation{%
  \institution{Federal University of Pará}
  \city{Belém}
  \state{Pará}
  \country{Brazil}}
\email{gpinto@ufpa.br}

\renewcommand{\shortauthors}{Ferreira et al.}

\begin{abstract}


Machine learning is increasingly being embedded into government digital platforms, but public-sector constraints make it difficult to build ML systems that are accurate, auditable, and operationally sustainable. In practice, teams face not only technical issues like extreme class imbalance and data drift, but also organizational barriers such as bureaucratic data access, lack of versioned datasets, and incomplete governance over provenance and monitoring. Our study of the Brasil Participativo (BP) platform shows that common engineering choices — like using LLMs for pre-labeling, splitting models into routed classifiers, and generating synthetic data — can speed development but also introduce new traceability, reliability, and cost risks if not paired with disciplined data governance and human validation. This means that, in the public sector, responsible ML is not just a modeling problem but an institutional engineering problem, and ML pipelines must be treated as civic infrastructure.
Ultimately, this study shows that the success of machine learning in the public sector will depend less on breakthroughs in model accuracy and more on the ability of institutions to engineer transparent, reproducible, and accountable data infrastructures that citizens can trust.

\end{abstract}




\maketitle

\section{Introduction}


The growing adoption of machine learning components in software systems has introduced engineering challenges that go beyond model selection and accuracy optimization~\cite{huyen2022designing, amershi_software_2019}. Unlike conventional code components, ML-based systems are dynamic, data-dependent, and probabilistic by nature~\cite{alves2023practices, amershi_software_2019, serban2020adoption}, making it challenging to ensure reproducibility, traceability, and accountability throughout their lifecycle. Adaptations of traditional software engineering practices are used to support ML-based systems development~\cite{alves2023practices}, but there is no specific~\cite{Parida_2025}. Thus, AI engineering becomes a separate discipline~\cite{bosch2021engineering}, bringing new methods and principles to address the unique needs of ML-based systems~\cite{amershi_software_2019, serban2020adoption}.

On government digital services, operational and ethical risks arise when machine learning algorithms induce bias in citizen-related decisions~\cite{rychwalska2019data, ozili2025digital}. Bias and errors in ML-based systems in the public sector can violate citizen rights and weaken public trust, in addition to driving up operational costs~\cite{ozili2025digital}. Therefore, designing ML systems for public services is challenging because it must balance technical robustness with social values such as accountability, fairness, and transparency~\cite{rychwalska2019data, ozili2025digital}. 

Engineering practices such as MLOps and other data governance frameworks can help provide infrastructure to ML development~\cite{kreuzberger2023machine,politz2017}. Still, existing solutions often remain too generic or infrastructure-centric to address the specific demands of public-sector problems~\cite{devops_2022}, and they presuppose stable data pipelines, plentiful expert annotations, and well-developed automation. Such assumptions seldom hold in civic technology contexts~\cite{lewis2019componentmismatchescriticalbottleneck, OECD2025}. As a result, software engineering teams working in these domains must deal with non-standardized data, limited computational resources, and high expectations for transparency and explainability that traditional development methods cannot easily meet.

In this paper, we investigate the challenges involved in integrating machine learning components into a large-scale governmental digital participation platform. Specifically, we examine the deployment of ML modules within Brasil Participativo, Brazilian civic platform that enables citizens to directly contribute to the formulation and improvement of public policies. Our study is driven by two research questions: (RQ1) What technical, organizational, and governance challenges emerge when designing and operating ML systems in the public sector? (RQ2) How do architectural decisions — including LLM-based pre-labeling strategies, class balancing, and modularization — influence the reliability, traceability, and operational cost of ML pipelines in civic infrastructure? To address these questions, we employ a retrospective mixed-methods approach that combines document analysis, semi-structured interviews, sample-based revalidation, and controlled experiments.




Our results show that LLM-based pre-labeling accelerates development during periods of limited human availability but requires continuous expert validation to maintain trust; that synthetic resampling and data augmentation provide limited gains in long-tail taxonomies; and that dual-classifier architectures improve performance metrics at the cost of increased operational complexity. Together, these findings contribute with practical evidence and engineering recommendations for building reproducible, auditable, and sustainable ML pipelines, in particular, for public-sector systems. To facilitate replication, our artifacts are available for use in the following url: \url{ https://drive.google.com/drive/folders/1imQerH0_GW3Z9nfTCMjz36_rWUTZe72z?usp=sharing}.

\section{Background \& Related Work}

This section reviews the civic and technical foundations of the Brasil Participativo platform and summarizes prior research on the engineering challenges of building and maintaining ML-enabled systems.

\subsection{The Brasil Participativo Platform}

The Brasil Participativo (BP) platform was launched in 2023 by the General Secretariat of the Presidency of the Republic to enable citizens to directly contribute to the creation, monitoring, and improvement of public policies. Over its first two years of operation, the platform registered approximately 1.6 million users and over 42,000 text-based proposals submitted by citizens\footnote{\url{https://www.gov.br/planejamento/documentos-hospedados-para-gerar-qrcodes/relatorio-ppaparticipativo}}. The platform is based on Decidim — an open-source, monolithic framework developed in Ruby on Rails — which was technically adapted and customized for the Brazilian context~\cite{rsp2024}.

The Proposals component allows citizens to create, comment on, like, and vote on policy proposals. It is the most actively used participation modality within the platform, concentrating the highest levels of engagement and mobilization. However, evaluating these proposals individually poses a significant challenge, given the volume of more than 42,000 submissions and the limited availability of staff. In practice, only the most upvoted proposals tend to be formally assessed and incorporated into public policies. The adoption of machine learning techniques aims to facilitate the large-scale processing of citizen input and enhance participation dynamics. Automated proposal classification supports the identification of citizens’ interests based on their engagement patterns, enabling the design of personalized re-engagement and participation strategies.

\subsection{Engineering challenges in ML Systems}

\begin{figure}[t]
\centering
\resizebox{\columnwidth}{!}{
\begin{tikzpicture}[node distance=7.5mm]

\node[boxData] (p1) {\textbf{Problem definition:}\\ \scriptsize establishing objectives and collecting relevant data.};
\node[boxData, below=of p1] (p2) {\textbf{Data processing:}\\ \scriptsize cleaning, transforming, and preparing data for modeling.};

\node[handoff, below=8mm of p2] (hx)
{Handoff: Dataset / Feature Store\\ \scriptsize (contracts, documentation, SLAs)};

\node[boxML, below=12mm of hx] (p3) {\textbf{Model selection:}\\ \scriptsize training and evaluating candidate models according to desired metrics.};
\node[boxML, below=of p3] (p4) {\textbf{Software development:}\\ \scriptsize implementing ML and non-ML components of the system (e.g., user interface).};
\node[boxML, below=of p4] (p5) {\textbf{ML integration:}\\ \scriptsize incorporating the trained ML model into the overall system architecture.};
\node[boxML, below=of p5] (p6) {\textbf{System testing:}\\ \scriptsize validating that all components function correctly and meet end-to-end requirements.};
\node[boxML, below=of p6] (p7) {\textbf{Deployment, monitoring, and maintenance:}\\ \scriptsize deploying the system, continuously monitoring its performance, and addressing errors and model drift over time.};

\draw[flow] (p1) -- (p2);
\draw[flow] (p2) -- (hx);
\draw[flow] (hx) -- (p3);
\draw[flow] (p3) -- (p4);
\draw[flow] (p4) -- (p5);
\draw[flow] (p5) -- (p6);
\draw[flow] (p6) -- (p7);

\begin{scope}[on background layer]
  \node[laneBG=LaneA, fit=(p1)(p2)] (bgDE) {};
  \node[laneBG=LaneB, fit=(p3)(p7)] (bgML) {};
\end{scope}

\node[laneLabel, anchor=north west] at ([xshift=1mm,yshift=0.75mm]bgDE.north west) {Data Engineering};
\node[laneLabel, anchor=north west] at ([xshift=1mm,yshift=-1mm]bgML.north west) {ML Engineering};

\end{tikzpicture}
}
\caption{ML system development workflow~\cite{amershi_software_2019, giray2021softwareengineeringperspectiveengineering, paleyes2022challenges}}
\label{fig:ml-workflow}
\end{figure}
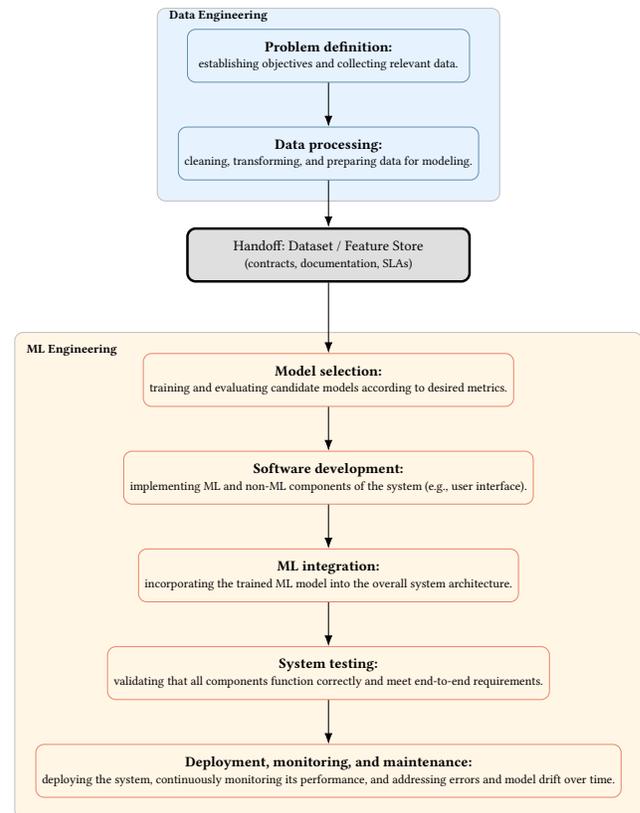

Traditional software is defined through explicit, predefined rules and instructions that deterministically process data to produce expected results~\cite{alves2023practices,arpteg2018software}. On the other hand, the ML systems derive their behavior from the data itself. The development of ML-based systems, therefore, inherently involves greater uncertainty and risk~\cite{alves2023practices, serban2020adoption, amershi_software_2019}, necessitating new engineering, management, and organizational approaches. Although methodologies such as Agile and Waterfall can be adapted to ML-enabled systems, there is still no universally accepted engineering process designed explicitly for ML~\cite{Parida_2025}. Most studies present a generalized ML development process, summarized by Parida et al.~\cite{Parida_2025} and further discussed in the literature~\cite{amershi_software_2019, giray2021softwareengineeringperspectiveengineering, paleyes2022challenges}, which is illustrated in Figure \ref{fig:ml-workflow}.

Arpteg et al.~\cite{serban2020adoption} identify 12 challenges spanning development, deployment, and organizational issues. Tasks such as managing and versioning data during development, monitoring and logging data for deployed models, and estimating the effort needed to develop ML components differ significantly from traditional software development.  Data-related challenges are particularly critical. ML systems heavily depend on large volumes of high-quality data. Issues such as data versioning, provenance tracking, and managing evolving datasets introduce significant complexity~\cite{schelter2017, politz2017}. Beyond technical aspects, ensuring data governance adds more activities to the development process. Polyzotis et al.~\cite{politz2017} and van der Weide et al.~\cite{vanderweide2017} highlight challenges and solutions for versioning and provenance in ML pipelines. Martínez-Fernández et al.~\cite{martinez2022} show that data-related issues remain the most recurrent challenge in AI engineering, making robust versioning and governance mechanisms central to the sustainability of ML systems.

Decision-making activities present additional challenges. Requirements analysis and specification remain critical bottlenecks in ML system development \cite{ishikawa2019engineers}. Unlike in traditional software engineering, where requirements are typically defined upfront and validated through acceptance testing at the end,  ML systems cannot rely on this sequential process, as it is impossible to guarantee or even estimate the model’s accuracy in advance. This challenge is echoed in recent surveys, which highlight the inherent difficulty of eliciting and validating requirements for ML-based systems under uncertainty~\cite{martinez2022}. According to Lwakatare et al.~\cite{lwakatare2019taxonomy}, the experimentation and prototyping stages are full of challenges, including vague problem formulations, poorly defined outcomes, and difficulties in establishing suitable baselines for evaluation.

According to Huyen \cite{huyen2022designing}, ML projects involve a set of roles. Data scientists and ML engineers, DevOps and platform engineers, and subject-matter experts (SMEs). SMEs provide different perspectives, which contribute essential domain knowledge that must be translated into versioned, testable artifacts. Still, there is often a lack of engineering backgrounds, making communication and tooling difficult. As Kostova et al.~\cite{serban2020adoption} pointed out, aligning stakeholder expectations and ethical considerations is an important step in specifying requirements for ML systems to guarantee that contributions can be integrated into a sustainable pipeline.

In ML-enabled systems, developers need to train and test models before they can measure performance. This presents as a challenge, as it makes it difficult to plan the project around predictable deliveries~\cite{arpteg2018software}. Additionally, Wan et al.~\cite{wan2019does} identify the challenge of maintaining the same performance during testing and in production. 

Some aspects of testing and verification are still lacking in ML-enabled systems. The oracle problem, lack of realistic data, and absence of standardized regression testing, while verification must also address fairness, robustness, and adversarial resilience~\cite{benbraiek2018, riccio2020, zhang2020}. However, current practices remain in their early stages, with most studies relying on simplified datasets and overlooking threats to validity~\cite{martinez2022}.

Systems based on ML should enable scalability and infrastructure as the core concerns. For large NLP models, massive computational resources are required. And this brings us to the need for GPUs and advanced orchestration \cite{lwakatare2019taxonomy}. Lwakatare et al. also note that massively scalable infrastructure requires specialized schedulers (e.g., SLURM, Ray, Airflow) and dedicated engineering teams. In addition, scalability in production encompasses stringent latency and throughput requirements, underscoring the fact that scale cannot be achieved solely by designing specific models but also spans the complete lifecycle of ML systems from training to deployment.

Model deployment is ultimately considered one of the more serious stages, introducing a number of hurdles that go far beyond simply putting a trained model into production. Deployed models hardly ever work in isolation; they need to integrate with complex pipelines for data ingestion, feature stores, monitoring tools, and downstream applications. Paleyes et al. \cite{paleyes2022challenges} underscore that integration often faces barriers posed by hidden technical debt (e.g., difficulties in reusing code and models and the accumulation of engineering anti-patterns). Not only that, but the successful deployment also requires the cooperation of data scientists and software engineers. Deployment also involves managing environment and dependency mismatches, packaging models into reliable serving infrastructures, and meeting robust delivery requirements associated with latency and throughput \ cite {arpteg2018software,lwakatare2019taxonomy}. Deployment triggers a cycle of ongoing model monitoring and updates to address concept drift and to meet business and ethical needs~\cite{sculley2015hidden}.

\noindent
Taken together, these studies reveal that while the technical foundations of ML engineering are increasingly well understood, little is known about how these challenges manifest in public-sector environments—where constraints on infrastructure, governance, and accountability fundamentally reshape how ML systems are designed, validated, and maintained.

\section{Research Questions}

Our work focus on two important yet overlooked research questions. 

\begin{itemize}
\item \textbf{RQ1} What technical, organizational, and governance challenges emerge when designing and operating ML systems in the public sector?
\end{itemize}

Machine learning systems in the public sector are developed and deployed under constraints that differ substantially from those in commercial or research settings. Beyond optimizing model accuracy, government ML teams must ensure compliance with principles of transparency, accountability, and fairness, while navigating bureaucratic processes, fragmented infrastructures, and limited resources. These conditions create a complex interplay between technical, organizational, and governance factors that remains underexplored in empirical software engineering.
This research question aims to characterize how these dimensions intersect in practice—how infrastructure maturity, access policies, data management practices, and institutional rules influence the reliability and evolution of ML systems. By identifying these challenges, we seek to inform both practitioners and policymakers about the sociotechnical conditions necessary for sustainable and trustworthy ML adoption in government platforms.

\begin{itemize}
\item \textbf{RQ2} How do architectural decisions influence the reliability, traceability, and operational cost of ML pipelines in civic infrastructure?
\end{itemize}

Engineering decisions in ML pipelines are rarely neutral: choices such as pre-labeling strategies, class-balancing techniques, or modular architectures can shape not only predictive performance but also auditability, reproducibility, and operational costs. In civic infrastructures, these trade-offs are amplified by scale, heterogeneous data, and the need for explainable systems that can be audited by non-technical stakeholders. This question investigates the concrete effects of architectural choices—particularly those involving LLM-based automation—on key software engineering attributes such as reliability, traceability, and maintainability. Understanding these effects enables teams to design ML architectures that balance technical efficiency with institutional requirements for accountability and transparency, advancing the broader agenda of responsible AI engineering in the public sector.

\section{Methodology}
This study adopts a retrospective, mixed-methods design~\cite{johnson2004mixed} to analyze the evolution of the lifecycle of an automated text classification system within the BP context. The goal is to understand how technical and organizational decisions related to automated labeling, architectural evolution, and experiment governance have influenced the system’s reliability, rework, and operational maturity.

The methodology is structured around five axes: (i) retrospective document analysis, (ii) sample revalidation, (iii) semi-structured interviews, (iv) training and validation, and (v) experiments performed. Each axis addresses a complementary dimension of the system’s evolution and its engineering decisions.

\subsection{Retrospective Document Analysis}
We conducted a retrospective document analysis to reconstruct the temporal sequence of decisions and modifications in the \emph{Brasil Participativo} classification pipeline. The procedure relied exclusively on artifacts recorded in the project repository and drive, using the \textit{Audit.xlsx} spreadsheet as the primary corpus.

The spreadsheet consolidates records from the main tables \textit{Issues}, \textit{Commits}, \textit{Notebooks}, \textit{Notes}, \textit{Presentations}, \textit{Interviews}, and \textit{Human Evaluation}. Each record was treated as a unit of evidence with the fields \textit{id}, \textit{title}, \textit{summary}, \textit{approach}, \textit{creation\_date}, \textit{source\_type}, \textit{source\_id}, and \textit{link}. In total, the spreadsheet contains 49 entries with valid dates between 7 January 2025 and 18 October 2025, all with active links and no duplicates in the composite key (source\_type, source\_id).

We began by consolidating four auxiliary tables, \textit{Timeline}, \textit{Anchor Artifact}, \textit{Phases}, and \textit{Anchor Event}. We then performed data and date normalization, as well as record summarization where applicable. Next, we unified the records from the main tables into the \textit{Timeline} and ordered them chronologically for sequential reading. This procedure made it possible to identify chains of technical precedence, such as preprocessing definitions, \textit{notebook} executions, and implementation \textit{commits}, and to cluster evidence around stable objectives, making explicit the links between decision, experimentation, and materialization.

In \textit{Anchor Event}, we gathered records classified as anchor events, such as \textit{issues}, \textit{commits}, and deliveries, which signal important decisions, adoption of techniques, or identified delivery milestones. From this set, we mapped the project phases across the lifecycle and recorded them in \textit{Phases}, as presented in Table~\ref{tab:phases_audit}. For each phase, we provided a descriptive name, the period with start and end dates, the count of associated items, and references to the anchor events that support each boundary.

\begin{table*}[t]
\centering
\caption{Development phases derived from the Timeline audit. Entry counts reflect items recorded between the listed dates; anchors are representative artifacts within each phase.}
\label{tab:phases_audit}
\small
\begin{tabular}{lcccc}
\hline
Phase & Start & End & Entries (n) & Anchor events \\
\hline
Onboarding & 2025-01-07 & 2025-01-31 & 3 & \#AE1 and \#AE2 \\
Experimentation with BERTopic & 2025-02-01 & 2025-03-26 & 12 & \#AE3, \#AE4, and \#AE5 \\
Unsupervised vs. Semi-supervised & 2025-04-24 & 2025-08-17 & 13 & \#AE6, \#AE7, \#AE8, \#AE9, \#AE10, and \#AE11 \\
Supervised and Deploymentt & 2025-08-18 & 2025-10-18 & 21 & \#AE12, \#AE13, \#AE14, \#AE15, \#AE16, \#AE17, and \#AE18 \\
\hline
\end{tabular}
\end{table*}

Finally, to facilitate navigation, we compiled in \textit{Anchor Artifact} the items referenced in the RQ1 \textit{Findings}. The table lists, for each artifact, the project phase, title, item type, primary function, and a brief description of the corresponding evidence.

\subsection{Sample Revalidation}
A sample-based revalidation was conducted on the training set to assess the practical utility of automated labeling by large language models (LLMs). This analysis compared the labels originally generated by the model with human annotations obtained from a stratified sample of 200 records, allowing estimation of the degree of alignment between automatic inference and human judgment.

The sample was built by a script that selected 200 examples stratified by category (\texttt{VCGE\_N1}) from the test file previously labeled by the gemma3:12b model in a local environment. The process produced a file containing the examples, traceability metadata, the random seed (42), class distribution, and the annotation instructions. Three identical copies were provided to human experts, with the automatic labels masked, ensuring independent and blind annotation relative to the model’s outputs.

Evaluators followed a standardized protocol describing every step of the process. The instructions required individual labeling in strict accordance with the provided taxonomy. In cases of ambiguity, overlapping categories, or lack of an appropriate match, the experts were instructed to use the markers \texttt{ADJ} or \texttt{OTHER} and to record observations in a dedicated field. Use of any automated tools was explicitly prohibited.

After annotation was completed, agreement among the human annotators was estimated using Fleiss’ kappa~\cite{fleiss1971measuring}, complemented by pairwise agreement measures with Cohen’s kappa~\cite{kohen1960coefficient}. For comparison with the model, a majority-vote criterion (two or more agreeing evaluators) was adopted as the reference (\emph{majority gold}) \cite{braun2024beg}. Agreement between the LLM and this reference was then assessed—again using Cohen’s kappa—along with calculations of the mismatch rate and overall accuracy.

This approach enabled quantification of the alignment between the model’s predictions and human consensus, providing an empirical measure of the reliability of the automatic labels. The sample-based revalidation proved essential for assessing the robustness of the inferences produced by the LLM, highlighting both its potential as a labeling support tool and the limitations arising from the absence of adjudication and ongoing human review.

\subsection{Interviews}

To supplement our understanding of the solution development lifecycle and to capture undocumented perspectives related to data provisioning, we conducted a few semi-structured interviews with members from different technical teams involved in the project. The goal was to identify both technical and organizational factors that influenced decision-making throughout the development process, but were not reflected in existing documentation.

Three interviews were conducted on October 10, 2025, by the second author. Participants were selected based on their direct involvement in data engineering and related development activities. 
Each session followed a semi-structured guide. The guide contained eleven questions organized into five thematic areas, balancing comparability across participants with flexibility to explore emerging topics. All interviews were recorded with informed consent, transcribed, and analyzed. The interview guide was structured as follows:

\vspace{0.2cm}
\noindent
\textbf{Context of Work}
\begin{enumerate}
\item What is your current role within the team?
\item What are your main responsibilities?
\end{enumerate}

\noindent
\textbf{Processes and Tools}
\begin{enumerate}
    \item How do you evaluate the current processes for data ingestion, transformation, and publication?
    \item Are these processes clear and well documented?
    \item Do current tools meet the team’s needs?
\end{enumerate}

\noindent
\textbf{Access to External Data}
\begin{enumerate}
    \item Have you faced difficulties in accessing data due to bureaucracy, authorization issues, or unavailability?
    \item How do such problems impact the team’s work?
\end{enumerate}

\noindent
\textbf{Technical Challenges}
\begin{enumerate}
    \item What are the main technical problems you encounter?
    \item Can you describe any recent situations that were particularly challenging?
\end{enumerate}

\noindent
\textbf{Collaboration}
\begin{enumerate}
    \item How is the interaction with other areas, such as Machine Learning and Product Engineering?
    \item Are there communication barriers that hinder work across teams?
\end{enumerate}

The resulting analysis adopted an exploratory thematic approach aimed at identifying undocumented process aspects and practical constraints affecting data workflows. These interviews are not intended to support statistical generalization but rather to enrich the qualitative understanding of the development ecosystem and inform subsequent recommendations.

\subsection{Training and Validation}

All models followed a common preprocessing pipeline. Texts were standardized, empty rows and duplicates were removed, Portuguese stopwords were filtered, and lemmatization was applied. Sentences were represented with sentence transformers, predominantly BERTimbau~\cite{souza2020bertimbau}. In the supervised experiments, labels were normalized to lowercase and encoded with \texttt{LabelEncoder}. Each execution recorded artifacts and metadata for auditability and reproducibility, with more complete consolidation in the supervised setting due to the larger number of techniques and tests conducted.

In the unsupervised setting, BERTopic~\cite{grootendorst2022bertopic} was trained over fixed embeddings. Different sentence encoders and target numbers of topics were compared, with repeated runs per configuration. Topic quality was evaluated using Weight Score~\cite{hott2023evaluating} beetween c\_npmi coherence and vocabulary diversity, and external adherence to the \texttt{VCGE\_N1} taxonomy was estimated a posteriori through clustering metrics after filtering out outliers. The experimental design and the comparisons between unsupervised and semi-supervised variants are discussed in detail~\cite{ferreira2025semanticclusteringcivicproposals}, which serves as the technical reference for these two settings.

In the semi-supervised setting, BERTopic was initialized with terms derived from the \texttt{VCGE\_N1} taxonomy~\cite{gama2018vocabulario}, employing a class-TFIDF with seed amplification. Training used the same embeddings with fixed hyperparameters for topic size and count and produced topic metadata for external analysis. To verify the unsupervised and semi-supervised results, we relied on the cited article and on artifacts identified in the documentary analysis, including notebooks, repositories, commits, issues, and presentations that corroborate the reported findings.

In the supervised setting, a hyperparameter search was performed over classifiers operating directly on the embeddings, comparing logistic regression, random forests, SVM, and XGBoost with stratified five-fold cross-validation and accuracy as the selection criterion \cite{sokolova2009systematic}. The chosen classifier was then trained and an inference-ready bundle was produced, comprising the \texttt{LabelEncoder}, a probability-enabled classifier, and a topic-to-label mapper for auxiliary analysis. Validation used a held-out test split and reported classifier accuracy and macro-F1, topic model alignment metrics, and the rates of outliers and unmapped predictions. These procedures provide the documented and consistent basis for the controlled experiments described next.

\subsection{Experiments}

This subsection describes the methodology used to evaluate the supervised approach, strengthen pipeline reproducibility through systematic metadata, and observe model behavior under a production-like setting. The pipeline was refactored with emphasis on traceability and artifact versioning, producing, at each execution, versioned records and files that enable deterministic re-runs and auditability. These records include derived datasets, configuration parameters, checkpoints, and inference logs, together with manifests identifying artifacts and the test set via a cryptographic hash.

The experimental flow comprised preprocessing, training, packaging, and publishing models as HTTP services at a prediction endpoint. Although the infrastructure supports hyperparameter search, in the runs reported here the corresponding field was retained and versioned without being executed, thereby preserving a consistent training protocol across configurations.

We compared four supervised configurations. The \emph{normal} configuration does not employ explicit class-balancing techniques. The \emph{SMOTE}~\cite{chawla2002smote} configuration applies feature-space over-sampling. The \emph{synthetic-generation} configuration uses a local LLM, \texttt{gemma3:12b}, to produce additional examples. The \emph{dual-classifier} configuration partitions the label space into majority and minority sets using a threshold of two hundred examples per class and, at inference time, routes each request between two models according to service logic, resulting in multiple passes when necessary.

Validation was performed in batch mode on a test file with two thousand proposals, with text in \texttt{proposal\_text} and labels in \texttt{VCGE\_N1}. The evaluation script iterates over the set in batches, sends requests to the endpoint, and logs for each call the HTTP response code, timestamp, request identifier when available, final label and probability, and the full probability distribution. Each configuration was run ten times with a base seed of forty-two and unit increments per execution, ensuring stable estimates under controlled randomness.

Evaluation covered predictive and operational metrics. For predictive quality we computed accuracy, macro precision, macro recall, and macro F1, both aggregated and per class with support, to reveal tail behaviors. Operationally we measured mean latency and 95th-percentile latency, failure rate, and the average number of inferences per request. In single-classifier variants, inference occurs in a single pass; in the dual-classifier, routing between models can increase the average number of inferences per request and, consequently, the observed latency.

\section{Results}

In this section we provide answers to our research questions.

\subsection{RQ1 — Key technical, organizational, and governance challenges}

The combination of retrospective document analysis, sample revalidation, and interviews in the BP context characterizes three vectors that shape the pipeline’s reliability and auditability. The paragraphs below present \textbf{findings} grounded in evidence from the project itself.

\subsubsection{Technical}

\begin{figure}[h]
\centering
\includegraphics[width=\linewidth]{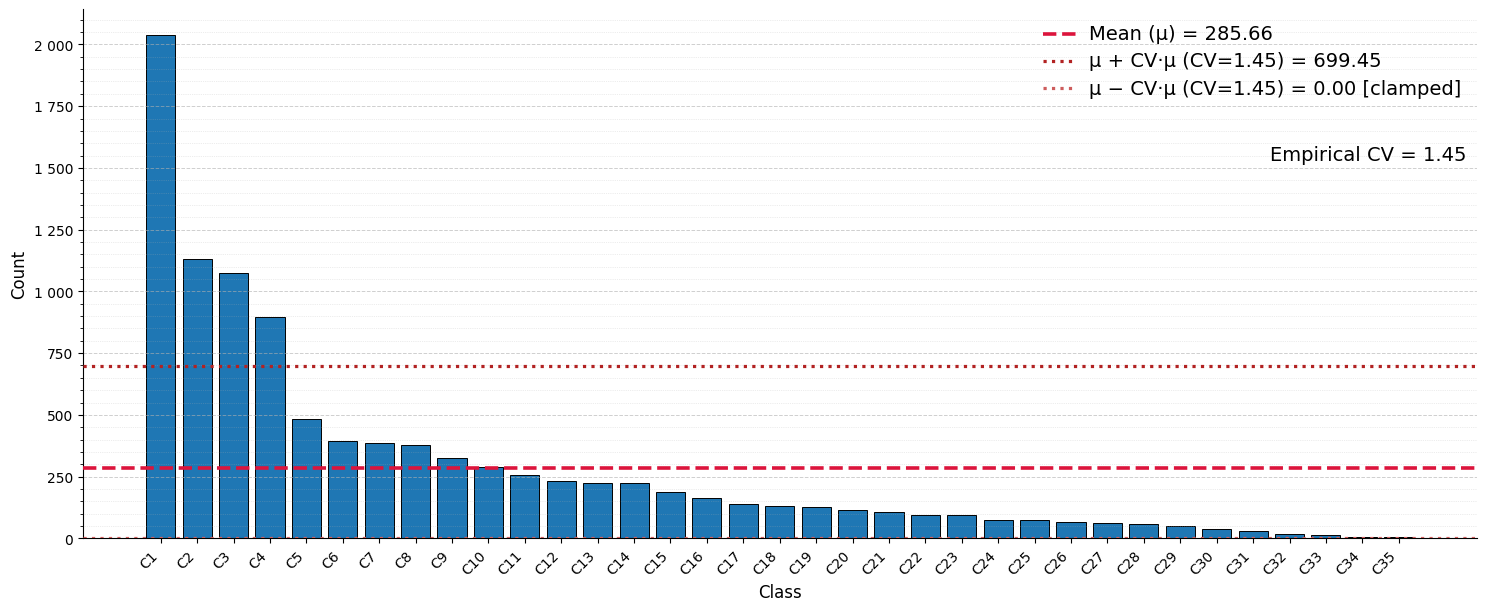}
\caption{Class distribution with mean and $\pm \mathrm{CV}\cdot u$ reference lines.}
\label{fig:class_distribution}
\end{figure}

\noindent\textbf{Finding T1: Long tail and per-class analysis.} The dataset is heavily imbalanced: there are 8,150 examples across 36 classes (mean $\approx 228$ per class), with the largest class containing 1,642 examples, the top three totaling 3,405 (41.8\%), and the three smallest summing to only 23. The largest–smallest ratio is 234:1 and the coefficient of variation is 1.45. Figure~\ref{fig:class_distribution} makes this skew clear: the x-axis shows classes \texttt{C1}\ldots\texttt{C36} and the y-axis shows the \textit{Count} of proposals per class; we mark the mean ($u=285{,}66$) with a dashed line and draw reference lines at $u+\mathrm{CV}\cdot u$ and $u-\mathrm{CV}\cdot u$ with $\mathrm{CV}=1{,}45$ (yielding $\approx 699{,}45$ and $\approx 0{,}00$, \textit{clamped}), along with the annotation “Empirical CV = 1.45.” Here, $u$ is the \textit{sample mean} of proposals per class and $\mathrm{CV}=\sigma/u$ is the \textit{empirical coefficient of variation} (standard deviation over the mean). This contrast between the mean and the $\pm \mathrm{CV}\cdot u$ band highlights the severe asymmetry of the dataset and motivates evaluating and optimizing performance by \textit{macro}-F1 and per-class F1, with emphasis on increasing recall for rare classes.

\noindent\textbf{Finding T2: Mitigating the long tail and preserving end-to-end quality under a routed topology.} We tested basic \textit{SMOTE}, synthetic example generation via LLMs, and a \textit{dual-classifier} with prevalence-based routing to ease the tail without altering the user-required taxonomy [\#AE16, \#C11, and \#NO6]. In training, the \textit{dual-classifier} looked most promising; however, in production-like simulation the overconfidence of the first stage kept items that should have proceeded to the second, lowering \textit{recall} on rare classes and depressing aggregate quality. We considered grouping rare classes into broader categories to reduce the tail, but this would sacrifice the granularity valued by analysts and political science teams, so we kept the taxonomy intact [\#NO7]. To ensure training gains carry through the real inference path, we began evaluating the pipeline \textit{end-to-end} in the \textit{dual-classifier} setting, monitoring the fraction of items that advance to the second stage, the share retained at the first, confidence distributions and calibration indicators, as well as the mean number of inferences per request and p95 latency. Based on these signals, the \textit{dual-classifier} may require threshold calibration and possibly a dedicated \textit{gating}/\textit{routing} model to curb the high confidence of the majority classifier; otherwise, promising training results do not translate into operational performance.

\noindent\textbf{Finding T3: LLMs under control and batch execution.} We integrated the LLM in a decoupled manner and processed it in controlled batches to accommodate the \textit{NVIDIA RTX 3090 with 24,GB of VRAM} [\#AE6, \#C1]; models up to 12B tend to saturate memory and reduce concurrency. We observed run-to-run variation even with fixed parameters; to maintain operational predictability, we adopted systematic \textit{logging} of inferences and timings. For classification, we recorded per item the \textit{timestamp}, model response, validated label, and duration, in addition to aggregate metadata. For synthetic generation, we saved the CSV of accepted examples and metadata including the generator model, \textit{temperature}, similarity threshold, maximum attempts, and \textit{runtime}, applying an \textit{embedding}-based similarity filter to avoid duplicates [\#C14]. In sum, the LLM delivered productivity gains under an \textit{offline}, monitored regime, with orchestration costs and concurrency limits proportional to model size, and with traceability sufficient for controlled re-executions.

\noindent\textbf{Finding T4: Human sampling to calibrate decisions.} As of October 2025, model evaluation has been predominantly automatic: weighting coherence and diversity in the unsupervised setting [\#AE8, \#P2]; ARI and NMI in the semi-supervised setting [\#AE9, \#N5, \#P3]; accuracy, F1, and \textit{macro}-F1 in the supervised setting [\#C9]. In light of T1 to T3, these signals alone are not sufficient for operational decisions. We conducted an initial round of human sampling (see RQ2) and sought to correlate judgments with WS, ARI/NMI, and classification metrics to define auditable acceptance criteria under the routed topology. \textit{Word}/\textit{topic intrusion} tasks also remain recommended to broaden interpretability assessments.

\subsubsection{Organizational}

\noindent\textbf{Finding O1: Static delivery without versioning, schema, or provenance.} Data was delivered primarily as a static file, with no versioned endpoint, no schema contract, and no provenance metadata [\#AE2, \#I1, \#N1, \#NO3]. This arrangement shifted the burden of reconstructing inputs to notebooks and scripts, hindered longitudinal comparisons, and increased the risk of silent divergences across development, validation, and production in BP. BP’s database grew substantially; between March and October 2025, 8,647 new proposals were recorded, yet there is no protocol for dataset updates  [\#I7]. There was only one update, from about 800 to approximately 10,000 proposals in the commit dated 2025-04-24 [\#C4], after which the corpus remained frozen, preventing the pipeline from reflecting potential \textit{data drift}. Operationally, dependence on a static \textit{drop} weakened the data pipeline: inputs had to be reconstructed manually, which compromised comparability across runs and widened the gap between environments. Without automated and integrated ingestion, temporal evaluations become infeasible, \textit{data drift} can go unnoticed, and models decouple from the production environment, reducing iteration speed, weakening traceability, and raising the risk of silent regressions.

\noindent\textbf{Finding O2: Bureaucracy to obtain Credentials and authorization to access data.} The bureaucracy of BP’s data access policy proved decisive for development cadence and the reliability of analytical artifacts. Although the Airflow, DBT, and Postgres services remained technically stable, dependence on government infrastructure teams for hosting and security, the use of a VPN, partial access grants, and rotating \textit{secrets} subjected the flow to mediated approvals and time-limited authorization windows, shifting the critical path from code readiness to credentials [\#IN1, \#IN2, \#IN3]. In practice, lead time came to depend on authorization queues, downstream materializations were postponed, and comparisons across runs lost robustness. To maintain cadence, teams resorted to local arrangements such as pre-labeling and \textit{ad hoc} sampling, which preserve speed but weaken central traceability and reinforce the static delivery pattern described in O1, making it harder to detect \textit{data drift} and to reproduce results. In terms of data maturity, BP is at an intermediate stage consistent with levels 1 to 2 of the DataOps Maturity Model, with manual deliveries, low automation, and informal processes. The effect is primarily organizational: the absence of a manageable and reproducible pipeline shifts effort to manual reconstruction and validation, reducing the time available for analysis and modeling.

\begin{quote}\small
``Our biggest obstacle is the bureaucracy to get access to any data or tool (\ldots) requests usually take two to three months, and initial database access took six to eight months.'' (E1)
\end{quote}

\noindent\textbf{Finding O3: No data catalog and no explicit rules.} Even with a central code repository on GitLab and the use of Metabase for querying and visualization, BP lacks a governed space that consolidates a data dictionary, business rules, accountable owners, and lineage across layers. In practice, metric definitions and transformation logic remain dispersed across notebooks and repositories, which produce variations in calculations and require manual reconstruction of criteria whenever results are compared. Metabase serves for exploration and \textit{dashboarding}, but it does not replace a versioned catalog with data contracts, \textit{owners}, change history, and explicit dependencies; nor does it provide a single, auditable register of rules applied outside the cards and \textit{queries} themselves. This fragmented arrangement hinders the reuse of variables for modeling, weakens end-to-end traceability and \textit{data drift} detection, and increases the cost of auditing and reproducibility. The organizational effect is clear: without a catalog and a systematic mapping of rules, cycle time comes to depend on tacit knowledge and \textit{ad hoc} validations, with a greater risk of inconsistencies across environments and decisions.

\subsubsection{Governance and auditability}

\textbf{Finding G1: Fragmented lineage and restricted re-execution.} In the analyzed period, traceability was concentrated in the code repository and analytical artifacts such as commits, notebooks, result tables, and synthetic generation files, but without a single chain linking file-delivered \textit{dataset cuts}, transformations, parameters, LLM prompts, \textit{checkpoints}, and inference logs through stable IDs [\#I6]. In practice, we reproduced runs deterministically when remaining within the same data \textit{drop} and the same code commit, supported by logs, \textit{seeds}, and local notes [\#N3, \#N5, \#N6, \#P1, \#P4]. When crossing different \textit{drops}, version comparisons required manual reconciliation of inputs and intermediate decisions and, in some cases, it was not possible to reconstruct a prior \textit{output} exactly. From a governance and auditability standpoint, the evidence trail exists but is fragmented by artifact and time period, which limits fine-grained differential analyses across versions and complicates retroactive investigations into regressions or model decisions.

\textbf{Finding G2: Limited per-version observability.} The production service is recent and, within the observed window, did not accumulate consolidated per-version time series for p95 and p99 latency, per-class error, or confidence distributions [\#AE14, \#I8]. Available monitoring remained closer to operational signals such as availability and logs, while analytical per-version readings derived mainly from experiments that mirrored system behavior in test environments. This arrangement allowed us to compare configurations, estimate latency costs, and observe per-class error variations, but without temporal continuity per version in production. From a governance standpoint, regression detection and explanation were feasible but anchored in experimental cycles [\#N3, \#N5, \#C1, \#C9, \#C11, \#P4] rather than operational per-version series, which increased the effort to assemble retrospective dashboards and to associate behavior changes with specific \textit{releases}, especially under a routed topology.

\textbf{Finding G3: Late and intermittent human sampling.} For much of the period, quality assessment centered on automatic metrics of coherence and diversity, ARI and NMI, and F1 and \textit{macro}-F1. The first structured round of human sampling occurred late [\#AE18] and revealed signals not captured by aggregate indicators, such as disagreements in rare classes, discrepancies between training performance and the real routed inference path, and cases where the confidence of the initial stage retained items that would have been better classified in the subsequent stage [\#HE2, \#HE3, \#HE4]. These judgments began to form qualitative evidence linked to versions, but still without a stable cadence and without a formal adjudication procedure when there was no majority. From a governance perspective, this expanded the auditability and contestability of results while also highlighting critical areas where human validation needs to become routine to sustain version-level acceptance criteria.

\subsection{RQ2 — Effects of LLM Pre-labeling, Class Imbalance Mitigation, and Architectural Decisions}

In this subsection, we report experimental and sample revalidation evidence that allows us to assess the trade-offs of three methodological interventions studied: (i) automated pre-labeling with LLMs, (ii) strategies to mitigate class imbalance through balancing or synthetic generation, and (iii) the architectural decisions adopted. For each intervention, we present reliability metrics (human agreement, per-class F1, macro-F1), traceability metrics (documentation and artifact generation), and operational cost metrics (latency, throughput, and implications for human review workload).

\subsubsection{LLM-based pre-labeling: reliability and traceability}
Human validation remains a critical pillar of reliability, and in this pipeline its scarcity was mitigated through automated pre-labeling with a language model. In supervised projects, expert labeling is a bottleneck that, in BP, was amplified by data volume and by texts that are non-standardized, ambiguous, colloquial, and thematically specific. In this cross-functional setting, the availability of annotators did not always align with the data science team’s demand, which motivated the use of LLMs as a temporary operational solution.

The technical implementation consisted of a Python script for batch labeling in a local environment. The corpus contained 10,186 records distributed across 36 classes. The first labeling of the \texttt{test.csv} file was completed on April 24, 2025 [\#T11] and recorded in the corresponding commit [\#C4]. The labeled set was kept as an operational reference for roughly six months, until October 2025, when we began sample audits to verify consistency and quality [\#AE18]. Pre-labeling was performed locally with the \texttt{gemma3:12b} model, processing 8,150 training records and 2,036 test records in approximately 94.6 minutes; files, logs, and metric tables were versioned to ensure traceability.

\begin{table*}
\centering
\caption{Summary of agreement and LLM accuracy for the audited sample (n=200)}
\label{tab:llm_prelabel_concise}
\begin{tabular}{lcc}
\hline
Metric & Value & Notes \\
\hline
Fleiss' kappa (3 annotators) & 0.6548 & 95\% CI 0.6125–0.6895; n=155 items with 3 labels \\
Pairwise kappas (p1–p2, p1–p3, p2–p3) & 0.8147, 0.3369, 0.3152 & n=166, 157, 172 \\
Items with majority label & 159/200 & 79.5\% \\
Items without majority label & 41/200 & 20.5\% \\
Complete disagreement & 13/155 & 8.4\% among items with 3 labels \\
\hline
LLM vs majority: Cohen's kappa & 0.481 & 95\% CI 0.400–0.557; n=159 \\
LLM vs majority: Accuracy & 50.314\% & n=159 \\
LLM vs majority: Mismatch rate & 49.686\% & 95\% CI 42.138\%–57.248\% \\
\hline
\end{tabular}
\end{table*}

Sample validation involved three experts independently labeling 200 stratified proposals. We defined \emph{majority gold} as agreement by at least 2 of 3 annotators. As summarized in Table~\ref{tab:llm_prelabel_concise}, 159 of 200 items (79.5\%) achieved a majority and composed the \emph{gold}; 41 items (20.5\%) did not reach a minimum consensus and were excluded from model comparisons. All model metrics were computed exclusively on the \emph{majority gold} (n=159); the 41 non-consensus items remain for adjudication in future work, and we report confidence intervals to quantify uncertainty. Among the 155 items with three valid labels, there were 13 cases of complete disagreement (8.4\%), indicating intrinsically ambiguous statements that impose a practical ceiling on observed performance.

Aggregate human agreement was substantial: Fleiss’ kappa of 0.6548, with a 95\% CI from 0.6125 to 0.6895, and asymmetry across pairs (p1 vs p2 = 0.8147; p1 vs p3 = 0.3369; p2 vs p3 = 0.3152). On the subset with a majority (n=159), the LLM achieved a Cohen’s kappa of 0.481 (95\% CI 0.400 to 0.557), accuracy of 50.314\%, and \emph{mismatch} of 49.686\% (95\% CI 42.138\% to 57.248\%). Taken together, the results indicate that the model reproduces part of the human pattern but disagrees in approximately half of the audited cases.

In the BP context, LLM pre-labeling accelerates coverage and produces reproducible operational labels, but sustained quality depends on review and adjudication. The presence of 20.5\% without a majority and 8.4\% complete disagreement recommends routine adjudication of these items and targeted audits in low-support classes, consolidating a more stable \emph{gold} and increasing the reliability of supervised training and monitoring.

\subsubsection{Balancing Strategies (SMOTE, LLM Synthetic, and Dual-Classifier)}

In the supervised training stage, we recorded complete metadata on training and test sizes, number of labels, classifier, embedding model, and metrics, and made them available in the documentary analysis artifacts. Table~\ref{tab:results_train_full} summarizes this offline scenario and shows that the \textit{Dual-Classifier} achieved the best aggregate accuracy and macro-F1 across distinct passes. These results initially supported adopting it, even with the expected operational cost of maintaining two inference paths.

\begin{table*}[t]
\centering
\caption{Comparison of models in training.}
\label{tab:results_train_full}
\small
\setlength{\tabcolsep}{3pt} 
\begin{tabular}{l l l r r r r r r r r r}
\hline
Model & Pass & Classifier & Train docs & Test docs & Num labels & Accuracy & Macro-F1 & ARI & NMI & V-measure & Duration (s) \\
\hline
DUAL-CLASSIFIER & pass1 & LogisticRegression & 5310 & 1524 & 11 & 0,7697 & 0,7491 & 0,5482 & 0,6013 & 0,6013 & 0,5129\\
DUAL-CLASSIFIER & pass2 & LogisticRegression & 1835 & 467  & 24 & 0,7537 & 0,6389 & 0,5632 & 0,7011 & 0,7011 & 0,2249\\
LLM              & pass1 & LogisticRegression & 11530 & 2000 & 35 & 0,6750 & 0,5296 & 0,4530 & 0,5667 & 0,5667 & 0,1905\\
NORMAL           & pass1 & LogisticRegression & 7872  & 1991 & 35 & 0,6781 & 0,5145 & 0,4522 & 0,5674 & 0,5674 & 0,3628\\
SMOTE            & pass1 & XGBClassifier      & 10674 & 2000 & 35 & 0,6165 & 0,4649 & 0,3719 & 0,4985 & 0,4985 & 0,4969\\
\hline
\end{tabular}
\end{table*}

For verifiability, we conducted experiments that simulate the production environment, aggregating per configuration macro-F1, accuracy, mean latency, and p95 latency over ten runs (Table~\ref{tab:results_agg}). The main findings are as follows. Synthetic augmentation with a local LLM offers the best balance between quality and cost, with macro-F1 of 0.5057 and accuracy of 0.6665, mean latency of 723.29 ms, and p95 of 1,005.70 ms. The unbalanced baseline trails slightly in quality, with macro-F1 of 0.4918 and accuracy of 0.6610, and shows essentially equivalent latency, with a mean of 725.00 ms and p95 of 1,007.70 ms. SMOTE modestly reduces latency, with a mean of 712.51 ms and p95 of 982.70 ms, but at a marked loss in quality, with macro-F1 of 0.4072 and accuracy of 0.6080. The \textit{Dual-Classifier} delivers the weakest aggregate performance, with macro-F1 of 0.3982 and accuracy of 0.5335, and the highest operational cost, with mean latency of 1,090.33 ms and p95 of 1,543.50 ms—consistent with routing between two models at inference.

\begin{table*}[t]
\centering
\caption{Aggregated results (mean over 10 runs) per configuration. Test set with 2000 proposals.}
\label{tab:results_agg}

\begin{tabular}{lcccc}
\hline
Model & Macro-F1 & Accuracy & Mean latency (ms) & P95 latency (ms) \\
\hline
LLM    & 0.5057 & 0.6665 &  723.29 & 1005.70 \\
NORMAL & 0.4918 & 0.6610 &  725.00 & 1007.70 \\
SMOTE  & 0.4072 & 0.6080 &  712.51 &  982.70 \\
DUAL-CLASSIFIER   & 0.3982 & 0.5335 & 1090.33 & 1543.50 \\
\hline
\end{tabular}
\end{table*}

The divergence between the \textit{Dual-Classifier}’s strong offline results and its degradation in the experiment stems from the training protocol and the absence of explicit confidence evaluation. The dataset was partitioned by prevalence, with majority classes in \textit{train\_part1.csv} and minority classes in \textit{train\_part2.csv}, and the test set split correspondingly into \textit{test\_part1.csv} and \textit{test\_part2.csv}. Offline, the passes were measured separately, which captures accuracy and macro-F1 but not confidence behavior. In production, the \textit{Dual-Classifier} employs a \textit{router}: inference first goes through \textit{pass1}, which proved overly confident even on examples within \textit{pass2}’s scope, skewing routing decisions and degrading the aggregate result. This effect does not occur in the NORMAL, SMOTE, and LLM variants, which do not employ inter-model routing.

In summary, the experiments favor the configuration with LLM-based synthetic data as the default for subsequent stages, while retaining the baseline as an operational reference. Under the evaluated parameterizations, SMOTE and the \textit{Dual-Classifier} are not recommended for this corpus. If the \textit{Dual-Classifier} is reconsidered, evaluation should be aligned with the production scenario and include confidence measurement and calibration before setting the decision threshold, mitigating \textit{pass1}’s excessive confidence over \textit{pass2}.

\section{Discussion}

Integrating ML components into large-scale civic platforms is a pressing challenge for both engineering and public-sector stakeholders. Models cannot be evaluated in isolation because their performance depends on processes that evolve over time, such as data delivery, annotation, and architectural definitions. Also, pipelines require continuous validation, monitoring, and redeployment. In this section, we summarize the main technical, organizational, and governance findings of the \emph{Brasil Participativo} case study, discuss their implications, and highlight opportunities for improvement and generalization.

\subsection{Technical}

From a technical standpoint, the study revealed a set of interdependent challenges typical of civic ML systems operating under limited resources. \textbf{Finding T1} showed that the dataset suffered from a severe long-tail imbalance (234:1 ratio), which undermined model recall for minority classes and emphasized the need for macro-F1 as a fairness-oriented metric. Attempts to mitigate this imbalance through \textbf{SMOTE} and LLM-based synthetic generation yielded mixed results—while LLM augmentation (\textbf{Finding T2}) improved performance modestly, the dual-classifier architecture underperformed in production due to routing overconfidence and latency overhead. Moreover, \textbf{Finding T3} demonstrated that LLM pre-labeling accelerated annotation but required strict control, logging, and batch execution to ensure reproducibility. These findings suggest that in resource-constrained public environments, improving pipeline reliability depends less on model complexity and more on systematic traceability, data versioning, and reproducible experimentation.

\subsection{Organizational}

Organizationally, the evidence points to structural bottlenecks that restrict engineering efficiency and data maturity. \textbf{Finding O1} identified the absence of a versioned data delivery mechanism, which forced teams to operate from static datasets and hindered drift detection. \textbf{Finding O2} highlighted that bureaucratic authorization processes and rotating credentials introduced delays, shifting the project’s critical path from code readiness to access management. \textbf{Finding O3} showed the lack of a unified data catalog, resulting in inconsistent metrics and duplicated logic across notebooks and dashboards. Collectively, these findings reveal that operational success depends as much on institutional agility and governance as on technical skill, reinforcing that sustainable ML engineering in the public sector requires investment in data infrastructure, documentation, and cross-team communication.

\subsection{Governance}

Governance-related findings underline the fragility of auditability in current civic ML deployments. The absence of end-to-end lineage (\textbf{Finding G1}), limited monitoring coverage (\textbf{G2}), and late human sampling (\textbf{G3}) constrain accountability and make regression analysis difficult. Without consistent evidence trails, silent data shifts or configuration errors can remain undetected, compromising trust and fairness in model-driven decisions. Because public-sector systems must adhere to principles of legitimacy and transparency, reproducibility and traceability are not optional—they are prerequisites for responsible AI. Strengthening audit trails, maintaining per-version metrics, and institutionalizing human validation cycles are thus essential steps toward ensuring that civic ML systems remain explainable, contestable, and aligned with public values.

\section{Limitations}

This study is subject to several limitations that should be considered when interpreting its findings. We group these limitations into methodological, contextual, and reproducibility aspects.

\textbf{Methodological limitations.} Our findings rely on operational proxies such as accuracy, macro-F1, latency, and traceability artifacts to infer pipeline reliability and maturity. While these indicators are grounded in empirical evidence, they may not fully capture broader constructs such as governance or institutional accountability. Similarly, fairness-related observations (e.g., long-tail bias) were based on quantitative performance metrics rather than social or ethical evaluations of impact. The evaluation of LLM-based pre-labeling also presents limitations: the agreement between the model and human annotators (Fleiss’ k = 0.65) assumes that the majority label represents truth, even though 20.5\% of the items lacked consensus and no adjudication round was performed. Consequently, the estimated reliability of automated labeling may be optimistic.

\textbf{Contextual limitations.} The case analyzed—\emph{Brasil Participativo}—represents a specific instance of civic ML deployment in a national public-sector platform. The context is characterized by restricted data access, bureaucratic approval processes, and infrastructural constraints typical of early-stage data governance programs. As such, our findings may not generalize to private-sector or large-scale commercial ML systems with mature automation pipelines, nor to more regulated domains such as healthcare or finance. Architectural trade-offs, such as the underperformance of the dual-classifier in production, are influenced by local infrastructure (on-premise GPUs, limited concurrency) and may behave differently under cloud-managed conditions.

\textbf{Reproducibility limitations.} While all experiments were executed under fixed seeds, hardware, and versioned configurations, full external replication is restricted. Civic data confidentiality prevents public release of raw datasets, and access to internal infrastructure (VPNs, credentialed databases) cannot be shared. To enhance transparency, we provide aggregate statistics, configuration details, and versioned artifacts; however, external teams can only analytically reproduce our results, not rerun the full pipeline. Despite these constraints, our documentation, audit trail, and structured methodology enable meaningful comparison and adaptation of our procedures to similar public-sector ML contexts.

\section{Conclusion}

This study analyzed the engineering, organizational, and governance challenges of integrating machine learning components into \emph{Brasil Participativo}, a large-scale civic participation platform in the Brazilian public sector. By combining retrospective document analysis and semi-structured interviews, we demonstrated that technical limitations—such as long-tail imbalance and fragile data provenance—are closely intertwined with organizational barriers like restricted data access, fragmented documentation, and low process automation. Our findings reveal that LLM-based pre-labeling can accelerate development under resource constraints but requires continuous human validation to maintain trust and traceability. Likewise, architectural strategies such as dual-classifier routing and synthetic augmentation introduce trade-offs between quality and operational cost. Ultimately, we argue that responsible ML engineering in public-sector contexts demands not only technical excellence but also robust data governance, reproducibility mechanisms, and interdisciplinary collaboration. Strengthening these aspects is essential for transforming government ML systems into auditable, transparent, and trustworthy infrastructures for large-scale civic engagement.

\subsection{Future Work}

Future research could extend this study in three directions.
First, by conducting longitudinal evaluations of ML systems in civic platforms, tracking how data drift, retraining frequency, and institutional processes evolve over time.
Second, by exploring human-in-the-loop validation frameworks tailored for public-sector settings, integrating expert review and citizen feedback into automated labeling and auditing workflows.
Third, by developing governance-aware MLOps architectures that embed versioning, explainability, and accountability mechanisms directly into ML pipelines, aligning technical observability with legal and ethical compliance.
Together, these directions can deepen understanding of how engineering, governance, and civic values interact, ultimately guiding the creation of sustainable and trustworthy AI infrastructures for democratic participation.

\bibliographystyle{ACM-Reference-Format}
\bibliography{sample-base}

\end{document}